\documentclass[prd,a4paper,onecolumn,nofootinbib,preprintnumbers]{revtex4}
\usepackage[top=2.8cm, bottom=2.8cm, left=2.4cm, right=2.4cm]{geometry}
\usepackage[T1]{fontenc}     
\usepackage{amsmath,amssymb,lmodern} 
\usepackage{graphicx}
\usepackage{diagbox}
\usepackage{hyperref}
\usepackage{color}
\usepackage{lipsum}
\usepackage{bm}
\usepackage{url}

\usepackage{xcolor}
\definecolor{blue}{rgb}{0.19,0.64,0.54}
\definecolor{reddish}{rgb}{0.65, 0.2, 0.2}
\definecolor{red}{rgb}{0.7,0.3,0.3}
\definecolor{darkgreen}{rgb}{0.2,0.7,0.3}
\definecolor{darkblue}{rgb}{0.3,0.40,0.48}
\definecolor{gray}{rgb}{.8,.8,.8}

\newcommand{\mytitle}{A systematic procedure to build the beyond generalized Proca field theory}
\hypersetup{,
    unicode=false,          
    pdftoolbar=true,        
    pdfmenubar=true,        
    pdffitwindow=false,     
    pdfstartview={FitH},    
    pdftitle={\mytitle},    
    pdfauthor={A. Gallego Cadavid and Y. Rodríguez},     
    pdfsubject={Subject},   
    pdfcreator={Creator},   
    pdfproducer={Producer}, 
    pdfkeywords={keyword1} {key2} {key3}, 
    pdfnewwindow=true,      
    colorlinks=true,       
    linkcolor=red,          
    citecolor=darkblue,        
    filecolor=magenta,      
    urlcolor=darkblue,           
    linktocpage=true
}

\newcommand{\ml}{\mathcal{L}_}
\newcommand{\Ref}[1]{Ref.~\cite{#1}}
\newcommand{\bea}{\begin{eqnarray}}
\newcommand{\eea}{\end{eqnarray}}
\newcommand{\eqn}[1]{(\ref{#1})}
\newcommand{\eq}[1]{Eq.~(\ref{#1})}
\newcommand{\eqs}[1]{Eqs.~(\ref{#1})}

\begin{document}

\preprint{PI/UAN-2019-650FT}

\title{\mytitle}

\author{Alexander Gallego Cadavid}
\email{alexander.gallego@uv.cl}
\affiliation{Escuela  de  F\'isica,  Universidad  Industrial  de  Santander, \\ Ciudad Universitaria, Bucaramanga 680002, Colombia and \\Instituto de Física y Astronomía, Universidad de Valparaíso, \\ Avenida Gran Bretaña 1111, Valparaíso 2360102, Chile}

\author{Yeinzon Rodr\'iguez}
\email{yeinzon.rodriguez@uan.edu.co}
\affiliation{Centro de Investigaciones en Ciencias B\'asicas y Aplicadas, Universidad Antonio Nari\~no, \\ Cra 3 Este \# 47A-15, Bogot\'a D.C. 110231, Colombia}
\affiliation{Escuela  de  F\'isica,  Universidad  Industrial  de  Santander, \\ Ciudad  Universitaria,  Bucaramanga  680002,  Colombia}
\affiliation{Simons Associate at The Abdus Salam International Centre for Theoretical Physics, \\ Strada Costiera 11, I-34151, Trieste, Italy}


\begin{abstract}
To date, different alternative theories of gravity, although related, involving Proca fields have been proposed. Unfortunately, the procedure to obtain the relevant terms in some formulations has not been systematic enough or exhaustive, thus resulting in some missing terms or ambiguity in the process carried out. In this paper, we propose a systematic procedure to build the beyond generalized theory for a Proca field in four dimensions containing only the field itself and its first-order derivatives. We examine the validity of our procedure at the fourth level of the generalized Proca theory. In our approach, we employ all the possible Lorentz-invariant Lagrangian pieces made of the Proca field and its first-order derivatives, including those that violate parity, and find the relevant combination that propagates only three degrees of freedom and has healthy dynamics for the longitudinal mode.  The key step in our procedure is to retain the flat space-time divergences of the currents in the theory during the covariantization process. In the curved space-time theory, some of the retained terms are no longer current divergences so that they induce the new terms that identify the beyond generalized Proca field theory. The procedure constitutes a systematic method to build general theories for multiple vector fields with or without internal symmetries.
\end{abstract}


\maketitle

\section{Introduction}
Einstein's theory of General Relativity is currently the most compelling and simplified theory of classical gravity. It has survived stringent tests on its validity in different scenarios: the expansion of the universe, the propagation of gravitational waves, the formation of the large-scale structure, as well as the strong gravitational field scenarios of neutron stars and black holes~\cite{acceleration1,Perlmutter:1998np,Ata:2017dya, Bennett:2012zja, Akrami:2018vks,Abbott:2016blz,Schimd:2004nq, Jain:2010ka,  Zhao:2011te, Clifton:2011jh, Koyama:2015vza,Ezquiaga:2018btd,  Ishak:2018his}. Despite its success, General Relativity is still considered as incomplete since any attempt to produce a quantum theory of gravity (see e.g. Refs. \cite{ Heisenberg:2018vsk, Agullo:2015qqa, Agullo:2013ai,Rovelli:2004tv}) has shown not to be satisfactory enough. Moreover, when its predictions are compared with cosmological observations, some authors argue that there exist hints pointing to modifications of the theory~\cite{Heisenberg:2018vsk, Agullo:2015qqa, Agullo:2013ai, Koyama:2015vza, Clifton:2011jh, Ishak:2018his,Ezquiaga:2018btd}.

Recently, a plethora of modified gravity theories have been proposed in order to avoid the assumption of 
two unknowns constituents of the Standard Cosmological Model (also called $\Lambda$CDM), namely, Dark Matter and Dark Energy 
~\cite{Weinberg:2008zzc,Ellisbook,Peter:2013avv,Amendola:2015ksp}. Although there exists a large amount of observational data to constrain most of these modified gravity theories, some of their sectors have only been partially explored, hence their full cosmological implications are still unknown~\cite{Koyama:2015vza, Clifton:2011jh, Ishak:2018his, Heisenberg:2018vsk,Ezquiaga:2018btd}. The general scheme in the formulation of these theories is the fulfilment of diffeomorphism invariance, unitarity, locality, and the presence of a pseudo-Riemannian spacetime in the action of the theory~\cite{ Heisenberg:2018vsk}. Nonetheless, any attempt to modify General Relativity inevitably introduces new dynamical degrees of freedom which, depending on the type of modification, could be of scalar, vector or tensor nature. Unfortunately, such formulation could lead to instabilities or pathologies in the theory~\cite{Heisenberg:2018vsk, Woodard:2006nt, Woodard:2015zca}. A known pathology is the Ostrogradsky's instability~\cite{Ostrogradsky:1850fid,Motohashi:2014opa, Woodard:2006nt,Woodard:2015zca, Rodriguez:2017ckc}, where the Hamiltonian is not bounded from below. The Ostrogradsky's theorem states that, \textit{for a non-degenerate theory}\footnote{A non-degenerate theory at $n$th-order is one in which its Lagrangian fulfils the condition $\det \left(\partial^2 \mathcal{L}/\partial q^{(n)}_i \partial q^{(n)}_j \right) \ne 0$, where $q^{(n)}_i$ is the $n$-th derivative of the generalized coordinate $q_i$ of the system.}, field equations higher than second order lead to an unbounded Hamiltonian from below~\cite{Ostrogradsky:1850fid, Motohashi:2014opa, Woodard:2006nt,Woodard:2015zca}. Thus, in order to formulate a well-behaved fundamental theory, we must build the action in such a way that the field equations are, at most, second order.

Three relevant formulations of such modified gravity theories correspond to scalar-tensor, vector-tensor, and scalar-vector-tensor theories, or simply Horndeski, generalized Proca, and scalar-vector-tensor gravity theories respectively~\cite{Horndeski:1974wa,Kobayashi:2019hrl,Deffayet:2013lga, Heisenberg:2018vsk,Deffayet:2011gz,Deffayet:2009wt,Deffayet:2009mn,Nicolis:2008in,Rodriguez:2017ckc,Horndeski:1976gi,Heisenberg:2014rta,Tasinato:2014eka,Allys:2015sht,Allys:2016jaq,Jimenez:2016isa,Heisenberg:2018acv}. These theories satisfy the necessary, but not sufficient, requirement to be free from the instabilities or pathologies previously mentioned since the actions are built so that the field equations are second order. Nowadays, extended versions of Horndeski and generalized Proca theories have been proposed, namely, beyond Horndeski, extended scalar-tensor\footnote{Also called degenerate higher-order scalar-tensor theories (DHOST).}, beyond generalized Proca (BGP), and extended vector-tensor theories\footnote{Which, by the way, could be called degenerate higher-order vector-tensor theories (DHOVT).}~\cite{Gleyzes:2014dya,Gleyzes:2014qga,Achour:2016rkg,Crisostomi:2016tcp,Crisostomi:2017aim,Crisostomi:2016czh,Deffayet:2015qwa,Gao:2014fra,Langlois:2015cwa,Langlois:2015skt,Lin:2014jga,Zumalacarregui:2013pma,Motohashi:2016ftl,BenAchour:2016fzp,Heisenberg:2016eld,Kimura:2016rzw}. Following similar procedures as those used to build the generalized Proca theory, the authors in Refs.~\cite{Allys:2016kbq,Jimenez:2016upj} obtained a massive extension of a SU(2) gauge theory, i.e., the generalized SU(2) Proca theory. This theory is also called the non-Abelian vector Galileon theory since it considers a non-Abelian vector field $A_\mu^a$, with $a=1,2,3$, whose action is invariant under the SU(2) global symmetry group. 

So far the generalized Proca and non-Abelian Proca field theories have been applied extensively to different phenomenological scenarios, which include the construction of inflationary cosmological models~\cite{Emami:2016ldl, Maleknejad:2011jw, Maleknejad:2011sq, Nieto:2016gnp,Oliveros:2019zkl}, the analysis of de Sitter solutions relevant to dark energy models~\cite{Rodriguez:2017wkg}, the study of their cosmological implications in the presence of matter~\cite{ Kase:2018nwt,Heisenberg:2018mxx,Kase:2018iwp,DeFelice:2016yws, DeFelice:2016uil,Nakamura:2017dnf}, the analysis of the strong lensing and time delay effects around black holes~\cite{Rahman:2018fgy}, and the construction of static and spherically symmetric solutions for black holes and neutron stars~\cite{Heisenberg:2017hwb, Rahman:2018fgy,Babichev:2016rlq,Chagoya:2017fyl,Kase:2018owh,Kase:2017egk}.

Although some physical and mathematical motivations to build alternative theories of gravity involving a Proca field $A_\mu$ have been given \cite{Horndeski:1976gi,Heisenberg:2014rta,Tasinato:2014eka}, the formulations have not been performed in a systematic enough or exhaustive way (see however Refs. \cite{Rodriguez:2017ckc,ErrastiDiez:2019trb,ErrastiDiez:2019ttn}). The purpose of this paper is to show a systematic procedure to build the most general Proca theory $\ml{n+2}^{\rm P}$ in four dimensions, where $\ml{n+2}^{\rm P}$ denotes the Lagrangians containing
$n \ge 1 $ first-order derivatives of $A_\mu$~\cite{Rodriguez:2017ckc, Heisenberg:2014rta,Tasinato:2014eka,Allys:2015sht, Allys:2016jaq,Jimenez:2016isa}. As an exception to the rule, $\ml{2}^{\rm P}$ is defined as the Lagrangian consisting of an arbitrary function of the Faraday tensor $F_{\mu \nu} \equiv \partial_\mu A_\nu - \partial_\nu A_\mu$, its Hodge dual $\tilde{F}_{\mu \nu} \equiv \epsilon_{\mu \nu \rho \sigma} F^{\rho \sigma} /2$, where $\epsilon_{\mu \nu \rho \sigma}$ is the Levi-Civita tensor, and $A_\mu$ only. As we will show, the theory thus built is equivalent to the BGP theory since we are able to obtain the Lagrangian $\ml{4}^{\mbox{\tiny N}}$~\cite{Heisenberg:2016eld} that identifies it. 

In some stages, the procedure is similar to that of Ref.~\cite{Allys:2016kbq}. The difference in our case resides in retaining 
the total derivatives of the flat space-time currents. These derivatives lead to some relations among Lagrangian pieces which, in turn, are used to eliminate some of the pieces since total derivatives do not contribute to the field equations. However, as we will show below, the convariantized versions of these relations, in some cases, are no longer total derivatives so they induce new terms in the curved space-time theory, hence leading to different field equations for the Lagrangians involved.

The layout of the paper is the following.  In Section~\ref{gp}, we describe the general procedure to construct the most general Proca theory.  In Section \ref{cfsc}, we discuss the issue of the total derivatives in flat spacetime and show how these terms are no longer total derivatives, in general, when going to curved spacetime.   Then, in Section~\ref{ip}, we implement the procedure to obtain the $\ml4^{\rm P}$ terms; there we show how to obtain systematically the  $\ml4^{\rm P}$ terms in the BGP. The conclusions are presented in Section~\ref{Conclusions}. Throughout the paper we use the signature $\eta_{\mu\nu} = \rm{diag}\,\left( -, +,+,+\right)$ and set $A \cdot A \equiv A_\mu A^\mu$ and $\partial \cdot A \equiv \partial_\mu A^\mu$. We also define the generalized Kronecker delta as $\delta^{\mu_1 \dots  \mu_{n-p}}_{\nu_1 \dots \nu_{n-p}} \equiv \delta^{[ \mu_1}_{\nu_1} \dots \delta^{\mu_{n-p} ]}_{\nu_{n-p}} = \delta^{\mu_1}_{[ \nu_1} \dots \delta^{\mu_{n-p} }_{\nu_{n-p} ]}$ where the brackets mean unnormalized antisymmetrization.

\section{General procedure}\label{gp}
In this section we describe in detail the procedure to build the most general theory for a Proca field containing only its first-order derivatives. For most of the description here, we follow the first steps of the procedure described in Ref.~\cite{Allys:2016kbq} until the consideration of the 4-currents. The procedure is as follows.

\subsection{Test Lagrangians}
Write down all possible test Lagrangians in a flat spacetime using group theory. The Lorentz-invariant quantities are constructed out of the metric $g_{\mu\nu}$ and the Levi-Civita tensor $\epsilon_{\mu\nu\rho\sigma}$. In Table~\ref{table:nLs}, we show the number of Lorentz scalars that can be constructed with multiple copies of $A_\mu$~\cite{Allys:2016kbq}, whereas, in Table \ref{table:nders}, we show the number of Lorentz scalars that can be built for a given product of vector fields and vector field derivatives~\cite{Allys:2016kbq}. These tables are non exhaustive.

	\begin{table}[h]
		\begin{tabular}{|l|c|c|c|c|c|c|c|c|}
		\hline
		number of vector fields $A^{\mu} $ & 1 & 2 & 3 & 4 & 5 & 6  \\
		\hline
		number of Lorentz scalars & 0  & 1 & 0 & 4 & 0 & 25  \\
		\hline
		\end{tabular}
		\caption{Number of Lorentz scalars that can be constructed with multiple copies of $A_\mu$.}
		\label{table:nLs}
	\end{table}
	
		\begin{table}[h]
		\begin{tabular}{|c|c|c|c|}
		\hline
		\backslashbox{number of $\partial^\mu A^\nu$}{number of $A^\rho A^\sigma$} & 0 & 1 & 2 \\
		\hline
		1 & 1   & 2   & 2  \\
		\hline
		2 & 4   & 10  & 11 \\
		\hline
		3 & 7   & 30  &  \\
		\hline
		\end{tabular}
		\caption{Number of Lorentz scalars that can be built for a given product of vector fields and vector field derivatives.}
		\label{table:nders}
	\end{table}
	
It is worth stressing that, when doing the respective contractions, 
some Lorentz scalars could be identical to each other and thus the number of independent terms would be reduced. 

Using group theory in this way, we can assure that all possible terms are written down, and that they are linearly independent.

\subsection{Hessian Conditions}\label{Hcgp}
Impose the condition that only three degrees of freedom for the vector field propagate~\cite{Rodriguez:2017ckc,Heisenberg:2014rta,Tasinato:2014eka,Allys:2015sht,Allys:2016jaq,Jimenez:2016isa,Allys:2016kbq,Jimenez:2016upj}. In order to achieve this, we first write down a linear combination of the test Lagrangians in the form
\begin{equation}
	\ml{\rm test} = \sum_{i=1}^{n} x_i \ml{i} \,,
\end{equation}
where $n$ is the number of test Lagrangians and $x_i$ are constant parameters of the theory. We then calculate the primary Hessian of the test Lagrangian
\begin{equation}\label{Hessian}
\mathcal{H}^{\mu\nu }_{\ml{\rm test}} \equiv \frac{\partial^2 \mathcal{L}_{\rm test}}{\partial \dot A_{\mu} \partial \dot A_{\nu}} \,,
\end{equation}
where dots indicate derivatives with respect to time. In order to ensure the propagation of only three degrees of freedom, we impose the vanishing of the determinant of the primary Hessian matrix $\mathcal{H}^{\mu \nu}$~\cite{Rodriguez:2017ckc,Heisenberg:2014rta,Tasinato:2014eka,Allys:2015sht,Allys:2016jaq,Jimenez:2016isa,Allys:2016kbq,Jimenez:2016upj}. This will guarantee the existence of one primary constraint that will remove the undesired polarization for the vector field. This condition is equivalent to satisfying $\mathcal{H}^{\mu 0}=0$,
i.e.,
\begin{equation}\label{Hmu0}
\mathcal{H}^{\mu 0}_{\ml{\rm test}} = \sum_{i=1}^n \mathcal{H}^{\mu 0}_{\ml{i}} = x_1 \mathcal{H}^{\mu 0}_{\ml{1}} +x_2 \mathcal{H}^{\mu 0}_{\ml{2}}+ \dots + x_n \mathcal{H}^{\mu 0}_{\ml{n}} = 0 \,.
\end{equation}
Eq.~\eqn{Hmu0} gives a system of algebraic equations for the $x_i$ whose roots impose conditions on the test Lagrangian. For some test Lagrangians their corresponding $x_i$ parameters will be zero, thus eliminating undesirable degrees of freedom. In practice, to calculate the primary Hessian condition in Eq.~\eqn{Hmu0}, it turns out to be easier to separately compute the cases $\mu=0$ and $\mu=i$.

As shown in Refs. \cite{ErrastiDiez:2019trb,ErrastiDiez:2019ttn}, the vanishing of the determinant of the primary Hessian matrix is not enough to guarantee the propagation of the right number of degrees of freedom when multiple vector fields are present.  In this case, an additional condition must be satisfied, namely the vanishing of the secondary Hessian $(\tilde{\mathcal{H}}_{\alpha \beta})_{\mathcal{L}_{\rm test}}$:
\begin{equation}
(\tilde{\mathcal{H}}_{\alpha \beta})_{\ml{\rm test}} \equiv \frac{\partial^2 \mathcal{L}_{\rm test}}{\partial \dot A_0^\alpha \partial A_0^\beta} - \frac{\partial^2 \mathcal{L}_{\rm test}}{\partial \dot A_0^\beta \partial A_0^\alpha} = 0 \,,  \label{seccons} 
\end{equation}
where the indices $\alpha$ and $\beta$ denote the different vector fields involved.
Nonetheless, keep in mind that, when going to a curved spacetime, the Hessian conditions are not sufficient to get rid of the ghost and Laplacian instabilities that might be present in the theory~\cite{Heisenberg:2014rta, Heisenberg:2016eld, Allys:2015sht, Allys:2016kbq,Deffayet:2015qwa}. To this purpose, the positiveness of the kinetic matrix and squared propagation speeds of the perturbation modes must be imposed respectively (see, for instance, Refs. \cite{Gomez:2019tbj, Jimenez:2013qsa}).

\subsection{Constraints among the Test Lagrangians}\label{cbtlgp}
Find constraints among the test Lagrangians that involve contractions among the Faraday tensor $F_{\mu \nu}$, its Hodge dual $\tilde{F}_{\mu \nu}$, and $A_\mu$. To this end, it is handy to use the identity~\cite{Allys:2016jaq, Fleury:2014qfa, Allys:2016kbq}
\begin{equation}\label{identity}
A^{\mu \alpha} \tilde{B}_{\nu \alpha} + B^{\mu \alpha} \tilde{A}_{\nu
\alpha} = \frac{1}{2} (B^{ \alpha \beta} \tilde{A}_{\alpha \beta}
)\delta^{\mu}_{\nu} \,,
\end{equation}
valid for all antisymmetric tensors $A$ and $B$. In Section~\ref{ip}, we will use this identity to find one constraint, thus eliminating one of the test Lagrangians. In \Ref{Allys:2016kbq}, a non-Abelian version of this identity was used to eliminate two test Lagrangians.

\subsection{Flat Space-Time Currents in the Lagrangian}\label{fsclgp}
Identify the Lagrangians related by total derivatives of the currents. 
In the case of Lagrangians involving two vector-field derivatives, it is useful to use the antisymmetric properties of the generalized Kronecker delta in order to define currents of the form~\cite{Allys:2016kbq}
\begin{equation}\label{Jdelta}
J^{\mu}_{\delta} \equiv f(X) \delta^{\mu \mu_2}_{\nu_1 \nu_2} A^{\nu_1} \partial_{\mu_2} A^{\nu_2} \,,
\end{equation}
where $X \equiv -A^2/2$.
We can also use the properties of the Levi-Civita tensor and define the following type of currents~\cite{Allys:2016kbq}:
\begin{equation} \label{Jepsilon}
J^\mu_\epsilon \equiv f(X) \epsilon^{\mu \nu \rho \sigma} A_\nu (\partial_\rho A_\sigma) \,.    
\end{equation}
Finally, we can define currents involving a divergence-free tensor $D^{\mu\nu}$~\cite{Allys:2016kbq}:
\begin{equation}\label{JD}
J^{\mu}_D \equiv f(X) D^{\mu\nu} A_{\nu} \,.
\end{equation}

From Eqs. \eqn{Jdelta} - \eqn{JD} we can write algebraic expressions among the test Lagrangians and total derivatives of the 4-current vectors. In a flat spacetime, we would use these relations to eliminate one or several test Lagrangians in terms of others since they yield the same field equations. However, in general, when the derivatives of the flat space-time currents are covariantized, what in flat spacetime are total derivatives, in curved spacetime are not anymore,
so the test Lagrangians that yield the same field equations in flat spacetime do not yield the same field equations in curved spacetime. 

Since this part of the procedure constitutes the main difference with respect to the approach followed in Refs. \cite{Rodriguez:2017ckc,Heisenberg:2014rta,Allys:2015sht,Allys:2016kbq,Jimenez:2016upj}, we will devote Section~\ref{cfsc} to explain this issue further.

\subsection{Covariantization}\label{cnacgp}
Covariantize the resulting flat space-time theory. To this purpose, we could simply follow the minimal coupling principle in which we replace all partial derivatives with covariant ones. One must also include possible direct coupling terms between the vector field and the curvature tensors~\cite{Allys:2015sht}. This procedure has been extensively explained in Refs.~\cite{Tasinato:2014eka,Jimenez:2016isa,Hull:2015uwa,Allys:2015sht,Heisenberg:2014rta,deRham:2011by,Jimenez:2013qsa} where the authors propose contractions on all indices with divergence-free tensors built from curvature, such as the Einstein $G_{\mu \nu}$ and the double dual Riemann $L^{\alpha\beta\gamma\delta}= - \frac12 \epsilon^{\alpha\beta\mu\nu}\epsilon^{\gamma\delta
\rho\sigma}R_{\mu\nu\rho\sigma}$ tensors.

\subsection{Scalar Limit of the Theory}\label{sltgp}

From Horndeski theories we have learned that, when gravity is turned on, it could excite the temporal polarization of the vector field, introducing new propagating degrees of freedom~\cite{Allys:2016jaq,Jimenez:2016isa,Heisenberg:2018vsk,deRham:2011by, Jimenez:2013qsa, Heisenberg:2014rta,Tasinato:2014eka,Allys:2015sht, Heisenberg:2016eld,Allys:2016kbq}.  This is the reason why, as a final step, we must verify that the field equations for all physical degrees of freedom, i.e. scalar and vector modes, are at most second order. To this end, we split $A^\mu$ into the pure scalar and vector modes
\begin{equation}
A^\mu= \nabla^\mu \phi + \hat{A}^\mu \,,
\end{equation}
where $\phi$ is the Stückelberg field and $\hat{A}^\mu$ is the divergence-free contribution ($\nabla_\mu \hat{A}^{\mu}=0$). 

For a theory built out of first-order derivatives in the vector field, the pure vector sector of $A^\mu$ cannot lead to any derivative of order higher than two in the field equations. As for the scalar part, derivatives of order three or more could appear when covariantazing, which can be expressed in terms of derivatives of some curvature terms and be eliminated, in turn, by adding the appropriate counterterms (arriving then to the Horndeski or beyond Horndeski theories in curved spacetime);  such counterterms can easily be generalized to the Proca field by employing the Stückelberg trick.  Care must be taken also with the mixed pure scalar-pure vector sector, following an identical procedure as the one described lines before\footnote{This, indeed, is the origin of the counterterm in the $\mathcal{L}_6^{\rm P}$ piece of the generalized Proca action.}.

It is worth mentioning that some of the built Lagrangians vanish in the scalar limit, indicating that these interaction terms correspond to purely intrinsic vector modes \cite{Allys:2015sht}.

\section{Covariantization of flat space-time currents}\label{cfsc}
As we explained above, from Eqs. \eqn{Jdelta} - \eqn{JD}, it is possible to write algebraic expression among the test Lagrangians and the total derivatives of the 4-current vectors. These relations would then be used to eliminate one or several test Lagrangians in terms of the others since they yield the same field equations in flat spacetime. However, when the gravity is turned on, the flat space-time current derivatives now involve curved space-time current derivatives and other curvature terms that arise because second-order derivatives, being promoted now to space-time covariant derivatives, do not commute anymore.  Thus, the test Lagrangians in the relation do not yield the same field equations. For instance, as we will see more clearly in the implementation, if we have an expression of the form
\begin{equation}\label{fsc}
	\partial_\mu J^\mu =\ml i + \ml j \,,
\end{equation}
which allows us to remove $\mathcal{L}_j$ in favour of $\mathcal{L}_i$ or viceversa,
a similar relation holds when promoting this expression to curved spacetime: 
\begin{equation}\label{csc}
	\nabla_\mu J^\mu =\ml i + \ml j + \mathcal{F}(A^\mu,\nabla^\mu A^\nu) \,,
\end{equation}
where $\mathcal{F}$ is a function of the field and the space-time covariant field derivatives. 
Nonetheless, we can see from this expression that, in curved spacetime, the field equations for $\ml i$ and $\ml j$ will no longer be the same due to the presence of the function $\mathcal{F}$. Anyway, it could also be the case that $\mathcal{F}$ vanishes identically, or that it is a total derivative, such that we are allowed to replace one of the Lagrangians in terms of the other 
since their field equations are the same.

\section{Implementation of the procedure}  \label{ip}
In this section, we will implement the procedure described in Section~\ref{gp}
for the case of the $\ml4^{\rm P}$ Proca Lagrangian. 
Paying attention to the covariantization of the flat space-time currents, discussed in Section \ref{cfsc}, we will arrive to the BGP theory whose main characteristic is its equivalence to the beyond Horndeski theory in the scalar limit.

\subsection{Test Lagrangians}
We start by writing all possible test Lagrangians for $\ml4^{\rm P}$. According to Table~\ref{table:nders}, in the case of two vector field derivatives only, there exist four terms which turn out to be independent:
\begin{equation} \label{l4-1}
\begin{array}{l}
\mathcal{L}_1 = (\partial \cdot A)^2 \,, \\
\mathcal{L}_2 = (\partial_\mu A_\nu)(\partial^\mu A^\nu) \,, \\
\mathcal{L}_3 = (\partial_\mu A_\nu)(\partial^\nu A^\mu) \,, \\
\mathcal{L}_4 = \epsilon^{\mu \nu \rho \sigma} (\partial_\mu A_\nu)(\partial_\rho A_\sigma) \,.
\end{array}
\end{equation}
In contrast, for two vector fields and two vector field derivatives 
there exist ten terms, six of them being independent:
\begin{equation} \label{l4-2}
\begin{array}{l}
\ml5=  (\partial \cdot A) (\partial_\rho A_\sigma) A^\rho A^\sigma \,, \\
\ml6=  (\partial_\mu A_\nu)(\partial^\mu A_\sigma)A^\nu A^\sigma \,, \\
\ml7=  (\partial_\mu A_\nu)(\partial_\rho A^\mu)A^\nu A^\rho \,, \\
\ml8=  (\partial_\mu A_\nu)(\partial_\rho A^\nu)A^\mu A^\rho \,, \\
\ml9=  \epsilon^{\mu \rho \sigma \beta} A_\beta (\partial_\nu A_\mu) (\partial_\rho A_\sigma) A^\nu \,, \\
\ml{10}=  \epsilon^{\mu \rho \sigma \beta} A_\beta (\partial_\mu A_\nu) (\partial_\rho A_\sigma) A^\nu \,, \\
\end{array}
\end{equation}
and the other four just being the same terms of Eq. (\ref{l4-1}) multiplied by $A^2$.  Regarding the test Lagrangians formed with four vector fields and two vector field derivatives, there exist eleven terms, four of them being the same terms of Eq. (\ref{l4-1}) multiplied by $A^4$, other six being the same terms of Eq. (\ref{l4-2}) multiplied by $A^2$ and the other one being
\begin{equation}
\ml{11} = (A^\mu(\partial_\mu A_\nu)A^\nu)^2 \,. \label{l4-3}    
\end{equation}
We can continue looking for test Lagrangians that contract two vector field derivatives with an even number of vector fields higher than four.  However, since the number of space-time indices corresponding to the two vector field derivatives is already saturated when considering the contractions with four vector fields, all the possible test Lagrangians that involve more than four vector fields will be exactly the same as the ones in Eqs. (\ref{l4-1}) - (\ref{l4-3}) multiplied by some power of $A^2$. This leads us to conclude that {\it all the possible test Lagrangians that involve two vector field derivatives are expressed as the ones in Eqs. (\ref{l4-1})-(\ref{l4-3}) multiplied each one of them by an arbitrary function of $A^2$}.

\subsection{Hessian Conditions}\label{Hc}
Continuing with the procedure, we now write down the linear combination of the terms in Eqs.~\eqn{l4-1}-\eqn{l4-3}, each one of them multiplied by an arbitrary function of $A^2$, to form the test Lagrangian
\begin{equation} \label{test}
\mathcal{L}_{\text{test}}=\sum_{i=1}^{11} f_i(X) \mathcal{L}_i \,,
\end{equation}
where the $f_i(X)$ are the mentioned arbitrary functions (the constants $x_i$ have been absorbed into the $f_i$). It is convenient to calculate first the primary Hessians in Eq.~(\ref{Hessian}) associated with the various test Lagrangians\footnote{The arbitrary functions $f_i(X)$ act, for this purpose, as constants since the primary Hessian calculation involves only derivatives of the test Lagrangians with respect to first-order field derivatives.}
\begin{eqnarray}
\mathcal{H}^{\mu\nu }_{\ml1} &=& 2 g ^{0 \mu} g ^{0 \nu} \,, \nonumber \\
\mathcal{H}^{\mu\nu }_{\ml2} &=& -2g^{\mu\nu} \,, \nonumber \\
\mathcal{H}^{\mu\nu }_{\ml3} &=& 2 g ^{0 \mu} g ^{0 \nu} \,, \nonumber \\
\mathcal{H}^{\mu\nu }_{\ml4} &=& 0 \,, \nonumber \\
\mathcal{H}^{\mu\nu }_{\ml5} &=& A ^{0} A ^{\mu} g^{0\nu}+A ^{0} A ^{\nu} g^{0\mu} \,, \nonumber \\
\mathcal{H}^{\mu\nu }_{\ml{6}} &=& -2 A^\mu A^\nu \,, \\
\mathcal{H}^{\mu\nu }_{\ml{7}} &=& A ^{0} A ^{\mu} g^{0\nu}+A ^{0} A ^{\nu} g^{0\mu} \,, \nonumber \\
\mathcal{H}^{\mu\nu }_{\ml{8}} &=& 2 (A^0)^2 g^{\mu \nu} \,, \nonumber \\
\mathcal{H}^{\mu\nu }_{\ml{9}} &=& 0 \,, \nonumber \\
\mathcal{H}^{\mu\nu }_{\ml{10}} &=& 0 \,, \nonumber \\
\mathcal{H}^{\mu\nu }_{\ml{11}} &=& 2 (A^0)^2 A^\mu A^\nu \,. \nonumber
\end{eqnarray}
Then imposing the primary Hessian condition in Eq.~\eqn{Hmu0} and considering the cases $\mu=0$ and $\mu=i$ separately, we obtain 
\begin{align}
\mathcal{H}^{00} = 2\left(f_1 + f_2 + f_3 \right) - 2 \left( f_{5} + f_{6} + f_{7} + f_8  \right) ( A^{0})^2 + 2 f_{11} (A^0)^4 &= 0 \,, \\
\mathcal{H}^{0i} =  -\left( f_5 + 2 f_{6} + f_{7} \right) \left( A^{0} A^{i } \right) + 2f_{11} (A^0)^3 A^i &= 0 \,,
\end{align}
leading to four independent algebraic equations which we solve for
\begin{equation}\label{Hcparams}
f_1 = -f_2 - f_3 \,, \quad f_{5} = - 2 f_{6} - f_{7} \,, \quad f_{6} = f_{8} \,, \mbox{ and } \quad f_{11} = 0 \,.
\end{equation}



Thus, our test Lagrangian in Eq. (\ref{test}) becomes
\begin{eqnarray} \label{testH}
\ml{\rm test} &=& f_2(X) (\ml{2}-\ml{1}) + f_3(X) (\ml{3} - \ml{1}) + f_4(X) \ml{4} \nonumber \\
&& + f_6(X) (\ml{6} - 2\ml{5} + \ml{8}) + f_7(X) (\ml{7} - \ml{5}) \nonumber \\
&& + f_9(X) \ml{9} + f_{10}(X) \ml{10} \,.
\end{eqnarray}
It is worth emphasizing that the secondary Hessian constraint of Eq. (\ref{seccons}) is trivially satisfied in this case since just one vector field is being considered.

\subsection{Constraints among the Test Lagrangians}\label{cbtl}
In this section, we make use of the identity in Eq.~\eqn{identity}~\cite{Allys:2016jaq, Fleury:2014qfa, Allys:2016kbq} in order to simplify the test Lagrangians. Let us consider $A^{\mu \nu} = F^{\mu \nu}$ and $B^{\mu \nu} = \tilde F^{\mu \nu}$. For these tensors, we can write down the relation
\begin{equation}
F^{\mu\alpha}\tilde{F}_{\nu \alpha} A_{\mu}A^{\nu} =
\frac{1}{4} \left(A\cdot A\right) F^{\alpha\beta } \tilde{F}_{\alpha\beta} \,.
\end{equation}
Now, expanding this expression in terms of the Proca field $A_\mu$ and its first-order derivatives, we obtain the following identity relating the Lagrangians in Eq.~\eqref{l4-2}:
\begin{equation}\label{l109}
\mathcal{L}_{9}-\mathcal{L}_{10} = \frac{1}{2} \mathcal{L}_{4} (A \cdot A) \,,
\end{equation}
which is also valid in curved spacetime. Using this relation we obtain
\begin{equation}\label{l9l10l4}
f_9(X) \ml{9} = f_9(X) \ml{10} - \ml{4} \ X f_9(X) \,.
\end{equation}
Therefore, recognizing that 
\begin{equation}\label{l4F}
\ml{4} = \frac{1}{2} F^{\alpha \beta} \tilde{F}_{\alpha \beta} \,,    
\end{equation}
which means that it actually belongs to $\ml{2}^{\rm P}= \ml{2}^{\rm P}( A_\mu, F_{\mu \nu},\tilde{F}_{\mu \nu})$~\cite{Rodriguez:2017ckc,Heisenberg:2014rta,Tasinato:2014eka,Allys:2015sht,Allys:2016jaq,Jimenez:2016isa,Allys:2016kbq,Jimenez:2016upj}, we can now write $f_9(X) \ml{9}$ in terms of $f_9(X) \ml{10}$ and a Lagrangian belonging to $\ml{2}^{\rm P}$, thus allowing us to remove $f_9(X) \ml{9}$ and $\left[f_4(X) - X f_9(X) \right] \ml{4}$ from $\ml{4}^{\rm P}$.

Another constraint can be found by noticing that
\begin{equation}\label{l2l3F}
(\ml{2}-\ml{1})-(\ml{3}-\ml{1}) = \frac{1}{2} F_{\mu \nu} F^{\mu \nu} \,,    
\end{equation}
so that $f_3(X) (\ml{3}-\ml{1})$ can be removed in favour of $f_3(X) (\ml{2}-\ml{1})$ and a Lagrangian belonging to $\ml{2}^{\rm P}$ (which can also be removed).

Thus, our test Lagrangian of Eq. (\ref{testH}) becomes
\begin{eqnarray} \label{testC}
\ml{\rm test} &=& [f_2(X)+f_3(X)] (\ml{2}-\ml{1})  \nonumber \\
&& + f_6(X) (\ml{6} - 2\ml{5} + \ml{8}) + f_7(X) (\ml{7} - \ml{5}) \nonumber \\
&& + [f_9(X) + f_{10}(X)] \ml{10} \,.
\end{eqnarray}

\subsection{Flat Space-Time Currents in the Lagrangian}
This part of the implementation is crucial since, from the flat space-time currents, we can obtain interaction Lagrangians which, before being promoted to curved spacetime, would be discarded in other methods.

Let us consider the current defined in Eq. (\ref{Jdelta}):
\begin{eqnarray}
J^{\mu}_{\delta} &\equiv& f(X) \delta^{\mu \mu_2}_{\nu_1 \nu_2}  A^{\nu_1} \partial_{\mu_2} A^{\nu_2} \nonumber \\
&=& f(X) \left[A^\mu (\partial \cdot A) - A^\nu \partial_\nu A^\mu \right] \,,
\end{eqnarray}
whose total derivative results in
\bea\label{jd}
\partial_{\mu}J^{\mu}_{\delta} &=& f(X) \left[ (\partial \cdot A)^2 + A^\mu \partial_\mu (\partial \cdot A) - \partial_\mu A^\nu \partial_\nu A^\mu - A^\nu \partial_\mu \partial_\nu A^\mu \right] \nonumber \\
&& - f_X (X) A^\nu \partial_\mu A_\nu \left[ A^\mu (\partial \cdot A) - A^\rho \partial_\rho A^\mu \right] \nonumber \\
&=& - f(X) (\ml{3} - \ml{1}) + f_X(X) (\ml{7} - \ml{5}) \,,
\eea
where $f_X(X) \equiv \partial f(X)/\partial X$, and we have used
\begin{equation}\label{RicciA}
A^\mu [\partial_\mu,\partial_\nu] A^\nu \equiv A^{\mu} \partial_\mu \partial_\nu A^{\nu} - A^{\mu} \partial_\nu \partial_\mu A^{\nu} = 0 \,,
\end{equation}
since, in flat spacetime, the partial derivatives of the Proca field commute. As we said before, this part of the calculation is crucial since, in a curved spacetime, the covariant derivatives of the Proca field do not commute.  We see from Eq. (\ref{jd}) that a term of the form $f_7(X) (\ml{7} - \ml{5})$ can be fully removed from $\ml{4}^{\rm P}$, {\it only in flat spacetime}, since it gives the same field equations as a term of the form $ (\int f_7(X) \ dX) (\ml{3} - \ml{1})$:
\begin{equation} \label{depcons}
f_7(X)(\ml{7} - \ml{5}) = \left( \int f_7(X) \ dX \right) (\ml{3} - \ml{1})  + \partial_\mu J^\mu_\delta \,. \end{equation}
The actual expression, without removing the commutator of the partial derivatives is:
\bea \label{boom}
f_7(X)(\ml{7} - \ml{5}) = \left( \int f_7(X) \ dX \right) (\ml{3} - \ml{1} - A^\mu [\partial_\mu,\partial_\nu] A^\nu)  + \partial_\mu J^\mu_\delta \,, \nonumber \\
= \left( \int f_7(X) \ dX \right) (\ml{2} - \ml{1} - A^\mu [\partial_\mu,\partial_\nu] A^\nu)  + \partial_\mu J^\mu_\delta \,, \eea
where in the last line we have replaced $\ml3-\ml1$ in terms of $\ml2-\ml1$ plus a term belonging to $\ml2^{\rm P}$, using \eq{l2l3F}. The latter term has been removed.

A similar procedure follows when considering the current defined in Eq. (\ref{Jepsilon}):
\begin{eqnarray}
\partial_\mu J^\mu_\epsilon &=& f(X) \left[ \epsilon^{\mu \nu \rho \sigma} (\partial_\mu A_\nu) (\partial_\rho A_\sigma) + \epsilon^{\mu \nu \rho \sigma} A_\nu (\partial_\mu \partial_\rho A_\sigma) \right] \nonumber \\
&& - f_X(X) A^\alpha \partial_\mu A_\alpha \left[\epsilon^{\mu \nu \rho \sigma} A_\nu (\partial_\rho A_\sigma) \right] \nonumber \\
&=& f(X) \ml{4} - f_X(X) \ml{10} \,,
\end{eqnarray}
showing that $\left[f_{9}(X) + f_{10}(X)\right] \ml{10}$ can be removed from $\ml{4}^{\rm P}$ since it gives the same field equations as a term belonging to $\ml{2}^{\rm P}$:
\begin{equation} \label{dJepsilon}
\left[f_{9}(X) + f_{10}(X)\right] \ml{10} = \left(\int \left[ f_9(X) + f_{10}(X) \right]  \ dX \right) \ml{4} - \partial_\mu J^\mu_\epsilon \,.   
\end{equation}
This last formula is valid {\it even in curved spacetime} because the commutation of partial second-order derivatives has not been invoked.


Finally, using Eq.~\eqn{JD}, and by virtue of the divergence-free properties of $\tilde{F}_{\mu \nu}$, we introduce the current 
\begin{equation}
J^{\mu}_F \equiv f(X) \tilde F^{\mu\nu} A_{\nu} \,,
\end{equation}
but this is nothing else than $J^\mu_\epsilon$, which leads us to the same results of Eq. (\ref{dJepsilon}).

The test Lagrangian of Eq. (\ref{testC}), after removing the redundant pieces, looks like
\begin{eqnarray} \label{testD}
\ml{\rm test} &=& \left \{ f_2(X)+f_3(X) + \int f_7(X) \ dX \right \} (\ml{2}-\ml{1})  \nonumber \\
&& + f_6(X) (\ml{6} - 2\ml{5} + \ml{8}) - \left( \int f_7(X) \ dX \right) A^\mu [\partial_\mu,\partial_\nu] A^\nu \nonumber \\
&=& \left \{ f_2(X)+f_3(X) + \int \left[2f_6(X) + f_7(X) \right] \ dX \right \} (\ml{2}-\ml{1})  \nonumber \\
&& + f_6(X) (\ml{6} - 2\ml{7} + \ml{8}) - \left \{ \int \left[2f_6(X) + f_7(X) \right] \ dX \right \} A^\mu [\partial_\mu,\partial_\nu] A^\nu \,,
\end{eqnarray}
where we have used \eq{boom} in the last line. We now notice that
\begin{equation}
(\ml{6} - 2\ml{7} + \ml{8}) = A_\mu A_\nu F^\mu_{\;\; \alpha}  F^{\nu \alpha} \,,
\end{equation}
so that it can be removed in favour of a Lagrangian belonging to $\ml2^{\rm P}$. Thus, our test Lagrangian becomes
\begin{eqnarray} \label{testD2}
\ml{\rm test} &=& \left \{ f_2(X)+f_3(X) + \int \left[2f_6(X) + f_7(X) \right] \ dX \right \} (\ml{2}-\ml{1})  \nonumber \\
&& - \left \{ \int \left[2f_6(X) + f_7(X) \right] \ dX \right \} A^\mu [\partial_\mu,\partial_\nu] A^\nu \,.
\end{eqnarray} 

\subsection{Covariantization}  \label{cnac}
In this section, we will covariantize our theory and show that it contains the BGP theory of Ref.~\cite{Heisenberg:2016eld} at the level of the $\ml4^{\rm P}$ sector of the Proca theory. We will also show how the $\ml4^{\rm P}$ sector of the BGP theory is induced by promoting the flat space-time currents into currents in curved spacetime.

By promoting all the partial derivatives to covariant ones in the test Lagrangian of Eq. (\ref{testD2}), the different pieces that it is made of can now be written as
\begin{eqnarray}\label{newbasiscst}
\left\{ f_2(X) + f_3(X) + \int \left[2f_6(X) + f_7(X)\right] \ dX \right\} (\ml{2}-\ml{1}) &=& -F_4(X) \delta^{\mu_1 \mu_2}_{\nu_1 \nu_2}(\nabla_{\mu_1} A^{\nu_1})(\nabla^{\nu_2} A_{\mu_2}) \,, \nonumber \\
-\left\{ \int \left[ 2f_6(X) + f_7(X) \right] \ dX \right\} A^\mu (\nabla_\mu \nabla_\nu - \nabla_\nu \nabla_\mu) A^\nu &=& G_N(X) R_{\mu \nu} A^\mu A^\nu \,, 
\end{eqnarray}
where $R_{\mu \nu}$ is the Ricci tensor and $F_4(X)$ and $G_N(X)$ are arbitrary functions of $X$. Therefore, our test Lagrangian can be written as
\begin{equation} \label{testl-2}
\ml{\rm test} = -F_4(X) \delta^{\mu_1 \mu_2}_{\nu_1 \nu_2}(\nabla_{\mu_1} A^{\nu_1})(\nabla^{\nu_2} A_{\mu_2}) + G_N(X) R_{\mu \nu} A^\mu A^\nu \,.
\end{equation}
The existence of the second term in the previous expression had not been recognized before, in Refs.~\cite{Rodriguez:2017ckc,Heisenberg:2014rta,Tasinato:2014eka,Allys:2015sht,Jimenez:2016isa,Allys:2016kbq,Jimenez:2016upj}, because the covariantization was performed over the final flat space-time Lagrangian, i.e., the one obtained after removing all the equivalent terms up to four-current divergences.  Nobody had paid attention to the fact that new terms could be generated in curved spacetime, terms that simply vanish in flat spacetime.

\subsubsection{Beyond generalized Proca  theory}\label{bgpevtt}
We will now show that the theory composed of the Lagrangians in \eq{testl-2} is the usual generalized Proca theory, before adding the required counterterms, plus the new BGP terms. To this end, we write the Lagrangian for two fields and two field derivatives unveiled in Ref.~\cite{Heisenberg:2016eld}:
\begin{equation}\label{L4N}
	\ml4^{\mbox{\tiny N}} = f_4^{\rm N}(X) \delta_{\alpha_1 \alpha_2 \alpha_3 \gamma_4}^{\beta_1 \beta_2\beta_3\gamma_4} A^{\alpha_1}A_{\beta_1} \nabla^{\alpha_2}A_{\beta_2} \nabla^{\alpha_3}A_{\beta_3}\,.
\end{equation}
Using the properties of the generalized Kronecker delta function, \eq{L4N} can be written as
\begin{eqnarray} \label{L4Nequiv}
\ml4^{\mbox{\tiny N}} &=& f_4^{\rm N}(X) \left[-2 X \delta^{\mu_1 \mu_2}_{\nu_1 \nu_2} (\nabla_{\mu_1} A^{\nu_1})(\nabla_{\mu_2} A^{\nu_2}) + (\nabla^\mu A_\nu) (\nabla_\rho A_\mu) A^\nu A^\rho - (\nabla \cdot A)(\nabla_\mu A_\rho)A^\mu A^\rho \right] \nonumber \\
&=&  f_4^{\rm N}(X) \left[-2X(\ml{1}-\ml{3}) - (\ml{5} - \ml{7})\right] \nonumber \\
&=& \left[-2X f_4^{\rm N}(X)-\int f_4^{\rm N}(X) \ dX \right] (\ml{1} - \ml{3}) + \nabla_\mu J^\mu_\delta + \left(\int f_4^{\rm N}(X) \ dX \right) R_{\mu \nu} A^\mu A^\nu \nonumber \\
&=& \left[-2X f_4^{\rm N}(X)-\int f_4^{\rm N}(X) \ dX \right] \left[\frac{1}{2}F^{\mu \nu}F_{\mu \nu} - (\ml{2} - \ml{1})\right] + \nabla_\mu J^\mu_\delta + \left(\int f_4^{\rm N}(X) \ dX \right) R_{\mu \nu} A^\mu A^\nu \,, \nonumber \\
&&
\end{eqnarray}
where the covariantized versions of Eqs. (\ref{l2l3F}) and (\ref{boom}) have been used.  Therefore, after removing the total derivative and the term belonging to $\ml{2}^{\rm P}$, $\ml4^{\mbox{\tiny N}}$ turns out to be
\begin{equation} \label{L4NFinal}
\ml4^{\mbox{\tiny N}} = -\left[2XG_{N,X}(X)+G_N(X)\right]\delta^{\mu_1 \mu_2}_{\nu_1 \nu_2}(\nabla_{\mu_1} A^{\nu_1})(\nabla^{\nu_2} A_{\mu_2}) + G_N(X) R_{\mu \nu} A^\mu A^\nu \,,
\end{equation}
where
\begin{equation}
G_N(X) \equiv \int f_4^{\rm N} (X) \ dX \,.   
\end{equation}
Thus, comparing Eqs. (\ref{testl-2}), (\ref{L4N}), and (\ref{L4NFinal}), we may conclude that our theory is equivalent to the BGP theory in the case of the $\ml4^{\rm P}$ Proca sector: 
\begin{equation} \label{testl-3}
\ml{\rm test} = -G_{4,X}(X) \delta^{\mu_1 \mu_2}_{\nu_1 \nu_2}(\nabla_{\mu_1} A^{\nu_1})(\nabla^{\nu_2} A_{\mu_2}) + \ml4^{\mbox{\tiny N}} \,,
\end{equation}
where
\begin{equation}
G_{4,X} \equiv F_4(X) - 2XG_{N,X}(X) - G_N(X) \,.
\end{equation}

\subsection{Scalar Limit of the Theory}\label{slt}
We will now verify that the longitudinal mode $\phi$ of the Proca field yields the correct scalar-tensor theory. We will show that our theory reduces to the beyond Horndeski theory \cite{Gleyzes:2014dya,Gleyzes:2014qga} in the scalar limit $A_\mu \to \nabla_\mu \phi$. In order to show this, we first write the Horndeski $\ml4^{\rm H}$ and beyond Horndeski $\ml4^{\rm BH}$ Lagrangians given in Refs.~\cite{Gleyzes:2014dya,Gleyzes:2014qga}:
\begin{eqnarray}
{\mathcal  L}_4^H &=& G_4(\phi, X)R - G_{4, X}(\phi, X) \Bigr( (\Box\phi)^2-\phi_{\mu \nu} \phi^{\mu \nu} \Bigl)  \,,  \label{hl4} \\
{\mathcal L}_{4}^{BH} &=& f_4^{\rm N}(\phi,X) \epsilon^{\mu\nu\rho}_{\ \ \ \ \sigma}\, \epsilon^{\mu'\nu'\rho'\sigma}\phi_{\mu}\phi_{\mu'}\phi_{\nu\nu'}\phi_{\rho\rho'} \notag \\ 
&=& f_4^{\rm N}(\phi,\, X) \Big[
X \Bigr( (\Box\phi)^2-\phi_{\mu \nu} \phi^{\mu \nu} \Bigl)
+2 \phi_{\mu} \phi_{\nu} \left( \phi^{\mu\alpha} \phi_{\alpha}^\nu - \Box \phi\, \phi^{\mu\nu} \right) \Big]\,, \label{bhl4}
\end{eqnarray}
where $X\,\equiv\,-\nabla_\mu \phi\,\nabla^\mu \phi/2$, $\phi_\mu \equiv \nabla_\mu \phi$, $\phi_{\mu \nu} \equiv \nabla_{\mu} \nabla_{\nu} \phi$, $R$ is the Ricci scalar, and $G_4$ and $f_4^{\rm N}$ are arbitrary functions of $\phi$ and $X$.

In the scalar limit $A_\mu \to \nabla_\mu \phi$, our test Lagrangian in \eq{testl-3} takes the form
\begin{equation}\label{limitnewbasiscst}
\ml{\rm test} \rightarrow -  G_{4, X}(X) \Bigr( (\Box\phi)^2-\phi_{\mu \nu} \phi^{\mu \nu} \Bigl) + f_4^{\rm N}(X) \Big[
X \Bigr( (\Box\phi)^2-\phi_{\mu \nu} \phi^{\mu \nu} \Bigl)
+2 \phi_{\mu} \phi_{\nu} \left( \phi^{\mu\alpha} \phi_{\alpha}^\nu - \Box \phi\, \phi^{\mu\nu} \right) \Big] \,,
\end{equation}
such that it reduces to the Horndeski and beyond Horndeski theories in \eqs{hl4} and \eqn{bhl4} respectively, except for the term proportional to the Ricci scalar. This means that our final $\ml{4}^{\rm P}$ Lagrangian is our test Lagrangian in Eq. (\ref{testl-3}) supplemented with a term $G_4(X) R$:
\begin{equation}
\ml{4}^{\rm P} = G_4(X) R -G_{4,X}(X) \delta^{\mu_1 \mu_2}_{\nu_1 \nu_2}(\nabla_{\mu_1} A^{\nu_1})(\nabla^{\nu_2} A_{\mu_2}) + f_4^{\rm N}(X) \delta_{\alpha_1 \alpha_2 \alpha_3 \gamma_4}^{\beta_1 \beta_2\beta_3\gamma_4} A^{\alpha_1}A_{\beta_1} \nabla^{\alpha_2}A_{\beta_2} \nabla^{\alpha_3}A_{\beta_3} \,.
\end{equation}

\section{Conclusions}  \label{Conclusions}
The generalized Proca theory is the vector field version of the Horndeski theory and, as such, satisfies a necessary condition required to avoid the Ostrogradsky's instability.  The original way to build it \cite{Heisenberg:2014rta,Jimenez:2016isa} consisted in finding out all the possible contractions of first-order vector field derivatives with a couple of Levi-Civita tensors, it being an extrapolation of the method employed in the construction of the scalar Galileon action which, in turn, lies on a formal demonstration given in Ref. \cite{Deffayet:2011gz}.  This method is very appropriate for the vector field case \cite{Heisenberg:2014rta,Jimenez:2016isa,Jimenez:2016upj}, even for the BGP theory \cite{Heisenberg:2016eld}, but it is incomplete since it does not generate parity-violating terms that we know exist in the theory \cite{Rodriguez:2017ckc,Allys:2015sht,Allys:2016kbq,Allys:2016jaq}; a very similar and formally proved methodology, which does generate the parity-violating terms, has been recently presented in Refs. \cite{ErrastiDiez:2019trb,ErrastiDiez:2019ttn}.

A more lengthy procedure was followed in Refs. \cite{Allys:2015sht,Allys:2016kbq,Allys:2016jaq} with the advantage that all the terms, including those that violate parity, can be produced.  This procedure does not rely on unproved hypothesis and, therefore, becomes a trustworthy way of building the generalized Proca theory.

Despite the methodology employed, however, earlier attempts did not take into account that what are total derivatives in flat spacetime may no longer be total derivatives in curved spacetime.  Thus, a few terms were ignored that we, in this paper, have unveiled, finding out that they produce the BGP terms.

Before finishing, let us discuss a bit about what the BGP theory is.  The BGP theory is a {\it non-degenerate} theory built from first-order space-time derivatives of the vector field and the field itself.  As such, its field equations are second order so that it satisfies the necessary requirement to avoid the Ostrogradsky's instability.  It satisfies the conditions for the propagation of the right number of degrees of freedom, at least in flat spacetime, and reduces to the beyond Horndeski theory in the scalar limit.  However, although this scalar limit corresponds to a {\it degenerate} theory, the full vector version, as we mentioned above, is not.  Having followed a lengthy but exhaustive procedure to build the generalized Proca theory, there was no reason at all not to find the BGP theory.  This was not the case in earlier attempts but the BGP theory should be there, hidden in some way.  We have discovered in this work that, in fact, the BGP theory at the $\mathcal{L}_4^{\rm P}$ level was hidden in those terms that look as total derivatives in the Lagrangian but that only are in flat spacetime.  We are then in the position to conclude that the BGP theory at the levels of $\mathcal{L}_5^{\rm P}$ and $\mathcal{L}_6^{\rm P}$ can be obtained following the systematic procedure described in this paper.  The method 
can also be applied to extensions of the generalized Proca theory, such as the scalar-vector-tensor theory developed in Ref. \cite{Heisenberg:2018acv} or the generalized SU(2) Proca theory of Refs. \cite{Allys:2016kbq,Jimenez:2016upj}.  Indeed, the construction of the beyond generalized SU(2) Proca theory will be discussed in a forthcoming paper \cite{alexyeinzon2}.



\section*{Acknowledgments}
A.G.C. dedicates this work to his mother María Libia Cadavid Carmona who is fighting cancer. A.G.C. thanks L. Gabriel Gómez for useful comments on the manuscript. A.G.C. was supported by Programa de Estancias Postdoctorales VIE - UIS 2019000052 and Beca de Inicio Postdoctoral 2019 UV. This work was supported by the following grants: Colciencias-DAAD - 110278258747 RC-774-2017, VCTI - UAN - 2017239, DIEF de Ciencias - UIS - 2460, and Centro de Investigaciones - USTA - 1952392. Some calculations were cross-checked with the Mathematica package xAct (www.xact.es).

\bibliographystyle{h-physrev4.bst}
\bibliography{Bibliography,BibliographyGravity}

\begin{thebibliography}{10}

\bibitem{acceleration1}
Supernova Search Team, A.~G. Riess {\em et~al.},
\newblock Astron. J. {\bf 116}, 1009 (1998), arXiv:astro-ph/9805201.

\bibitem{Perlmutter:1998np}
Supernova Cosmology Project, S.~Perlmutter {\em et~al.},
\newblock Astrophys. J. {\bf 517}, 565 (1999), arXiv:astro-ph/9812133.

\bibitem{Ata:2017dya}
SDSS, M.~Ata {\em et~al.},
\newblock Mon. Not. Roy. Astron. Soc. {\bf 473}, 4773 (2018), arXiv:1705.06373.

\bibitem{Bennett:2012zja}
WMAP, C.~L. Bennett {\em et~al.},
\newblock Astrophys. J. Suppl. {\bf 208}, 20 (2013), arXiv:1212.5225.

\bibitem{Akrami:2018vks}
Planck, Y.~Akrami {\em et~al.},
\newblock (2018), arXiv:1807.06205.

\bibitem{Abbott:2016blz}
Virgo, LIGO Scientific, B.~P. Abbott {\em et~al.},
\newblock Phys. Rev. Lett. {\bf 116}, 061102 (2016), arXiv:1602.03837.

\bibitem{Schimd:2004nq}
C.~Schimd, J.-P. Uzan, and A.~Riazuelo,
\newblock Phys. Rev. {\bf D71}, 083512 (2005), arXiv:astro-ph/0412120.

\bibitem{Jain:2010ka}
B.~Jain and J.~Khoury,
\newblock Annals Phys. {\bf 325}, 1479 (2010), arXiv:1004.3294.

\bibitem{Zhao:2011te}
G.-B. Zhao {\em et~al.},
\newblock Phys. Rev. {\bf D85}, 123546 (2012), arXiv:1109.1846.

\bibitem{Clifton:2011jh}
T.~Clifton, P.~G. Ferreira, A.~Padilla, and C.~Skordis,
\newblock Phys. Rept. {\bf 513}, 1 (2012), arXiv:1106.2476.

\bibitem{Koyama:2015vza}
K.~Koyama,
\newblock Rept. Prog. Phys. {\bf 79}, 046902 (2016), arXiv:1504.04623.

\bibitem{Ezquiaga:2018btd}
J.~M. Ezquiaga and M.~Zumalacárregui,
\newblock Front. Astron. Space Sci. {\bf 5}, 44 (2018), arXiv:1807.09241.

\bibitem{Ishak:2018his}
M.~Ishak,
\newblock Living Rev. Rel. {\bf 22}, 1 (2019), arXiv:1806.10122.

\bibitem{Heisenberg:2018vsk}
L.~Heisenberg,
\newblock Phys. Rept. {\bf 796}, 1 (2019), arXiv:1807.01725.

\bibitem{Agullo:2015qqa}
I.~Agullo and A.~Ashtekar,
\newblock Phys. Rev. {\bf D91}, 124010 (2015), arXiv:1503.03407.

\bibitem{Agullo:2013ai}
I.~Agullo, A.~Ashtekar, and W.~Nelson,
\newblock Class. Quant. Grav. {\bf 30}, 085014 (2013), arXiv:1302.0254.

\bibitem{Rovelli:2004tv}
C.~Rovelli,
\newblock {\em {Quantum gravity, }}Cambridge Monographs on Mathematical Physics
  (Univ. Pr., Cambridge, UK, 2004).

\bibitem{Weinberg:2008zzc}
S.~Weinberg,
\newblock {\em {Cosmology}} (Univ. Pr., Oxford, UK, 2008).

\bibitem{Ellisbook}
G.~Ellis, R.~Maartens, and M.~MacCallum,
\newblock {\em {Relativistic Cosmology}} (Univ. Pr., Cambridge, UK, 2012).

\bibitem{Peter:2013avv}
P.~Peter and J.-P. Uzan,
\newblock {\em {Primordial Cosmology, }}Oxford Graduate Texts (Univ. Pr.,
  Oxford, UK, 2013).

\bibitem{Amendola:2015ksp}
L.~Amendola and S.~Tsujikawa,
\newblock {\em {Dark Energy}} (Univ. Pr., Cambridge, UK, 2015).

\bibitem{Woodard:2006nt}
R.~P. Woodard,
\newblock Lect. Notes Phys. {\bf 720}, 403 (2007), arXiv:astro-ph/0601672.

\bibitem{Woodard:2015zca}
R.~P. Woodard,
\newblock Scholarpedia {\bf 10}, 32243 (2015), arXiv:1506.02210.

\bibitem{Ostrogradsky:1850fid}
M.~Ostrogradsky,
\newblock Mem. Acad. St. Petersbourg {\bf 6}, 385 (1850).

\bibitem{Motohashi:2014opa}
H.~Motohashi and T.~Suyama,
\newblock Phys. Rev. {\bf D91}, 085009 (2015), arXiv:1411.3721.

\bibitem{Rodriguez:2017ckc}
Y.~Rodríguez and A.~A. Navarro,
\newblock J. Phys. Conf. Ser. {\bf 831}, 012004 (2017), arXiv:1703.01884.

\bibitem{Horndeski:1974wa}
G.~W. Horndeski,
\newblock Int. J. Theor. Phys. {\bf 10}, 363 (1974).

\bibitem{Kobayashi:2019hrl}
T.~Kobayashi,
\newblock Rept. Prog. Phys. {\bf 82}, 086901 (2019), arXiv:1901.07183.

\bibitem{Deffayet:2013lga}
C.~Deffayet and D.~A. Steer,
\newblock Class. Quant. Grav. {\bf 30}, 214006 (2013), arXiv:1307.2450.

\bibitem{Deffayet:2011gz}
C.~Deffayet, X.~Gao, D.~A. Steer, and G.~Zahariade,
\newblock Phys. Rev. {\bf D84}, 064039 (2011), arXiv:1103.3260.

\bibitem{Deffayet:2009wt}
C.~Deffayet, G.~Esposito-Farese, and A.~Vikman,
\newblock Phys. Rev. {\bf D79}, 084003 (2009), arXiv:0901.1314.

\bibitem{Deffayet:2009mn}
C.~Deffayet, S.~Deser, and G.~Esposito-Farese,
\newblock Phys. Rev. {\bf D80}, 064015 (2009), arXiv:0906.1967.

\bibitem{Nicolis:2008in}
A.~Nicolis, R.~Rattazzi, and E.~Trincherini,
\newblock Phys. Rev. {\bf D79}, 064036 (2009), arXiv:0811.2197.

\bibitem{Horndeski:1976gi}
G.~W. Horndeski,
\newblock J. Math. Phys. {\bf 17}, 1980 (1976).

\bibitem{Heisenberg:2014rta}
L.~Heisenberg,
\newblock JCAP {\bf 1405}, 015 (2014), arXiv:1402.7026.

\bibitem{Tasinato:2014eka}
G.~Tasinato,
\newblock JHEP {\bf 1404}, 067 (2014), arXiv:1402.6450.

\bibitem{Allys:2015sht}
E.~Allys, P.~Peter, and Y.~Rodríguez,
\newblock JCAP {\bf 1602}, 004 (2016), arXiv:1511.03101.

\bibitem{Allys:2016jaq}
E.~Allys, J.~P. Beltrán~Almeida, P.~Peter, and Y.~Rodríguez,
\newblock JCAP {\bf 1609}, 026 (2016), arXiv:1605.08355.

\bibitem{Jimenez:2016isa}
J.~Beltrán~Jiménez and L.~Heisenberg,
\newblock Phys. Lett. {\bf B757}, 405 (2016), arXiv:1602.03410.

\bibitem{Heisenberg:2018acv}
L.~Heisenberg,
\newblock JCAP {\bf 1810}, 054 (2018), arXiv:1801.01523.

\bibitem{Gleyzes:2014dya}
J.~Gleyzes, D.~Langlois, F.~Piazza, and F.~Vernizzi,
\newblock Phys. Rev. Lett. {\bf 114}, 211101 (2015), arXiv:1404.6495.

\bibitem{Gleyzes:2014qga}
J.~Gleyzes, D.~Langlois, F.~Piazza, and F.~Vernizzi,
\newblock JCAP {\bf 1502}, 018 (2015), arXiv:1408.1952.

\bibitem{Achour:2016rkg}
J.~Ben~Achour, D.~Langlois, and K.~Noui,
\newblock Phys. Rev. {\bf D93}, 124005 (2016), arXiv:1602.08398.

\bibitem{Crisostomi:2016tcp}
M.~Crisostomi, M.~Hull, K.~Koyama, and G.~Tasinato,
\newblock JCAP {\bf 1603}, 038 (2016), arXiv:1601.04658.

\bibitem{Crisostomi:2017aim}
M.~Crisostomi, R.~Klein, and D.~Roest,
\newblock JHEP {\bf 1706}, 124 (2017), arXiv:1703.01623.

\bibitem{Crisostomi:2016czh}
M.~Crisostomi, K.~Koyama, and G.~Tasinato,
\newblock JCAP {\bf 1604}, 044 (2016), arXiv:1602.03119.

\bibitem{Deffayet:2015qwa}
C.~Deffayet, G.~Esposito-Farese, and D.~A. Steer,
\newblock Phys. Rev. {\bf D92}, 084013 (2015), arXiv:1506.01974.

\bibitem{Gao:2014fra}
X.~Gao,
\newblock Phys. Rev. {\bf D90}, 104033 (2014), arXiv:1409.6708.

\bibitem{Langlois:2015cwa}
D.~Langlois and K.~Noui,
\newblock JCAP {\bf 1602}, 034 (2016), arXiv:1510.06930.

\bibitem{Langlois:2015skt}
D.~Langlois and K.~Noui,
\newblock JCAP {\bf 1607}, 016 (2016), arXiv:1512.06820.

\bibitem{Lin:2014jga}
C.~Lin, S.~Mukohyama, R.~Namba, and R.~Saitou,
\newblock JCAP {\bf 1410}, 071 (2014), arXiv:1408.0670.

\bibitem{Zumalacarregui:2013pma}
M.~Zumalacárregui and J.~García-Bellido,
\newblock Phys. Rev. {\bf D89}, 064046 (2014), arXiv:1308.4685.

\bibitem{Motohashi:2016ftl}
H.~Motohashi {\em et~al.},
\newblock JCAP {\bf 1607}, 033 (2016), arXiv:1603.09355.

\bibitem{BenAchour:2016fzp}
J.~Ben~Achour {\em et~al.},
\newblock JHEP {\bf 1612}, 100 (2016), arXiv:1608.08135.

\bibitem{Heisenberg:2016eld}
L.~Heisenberg, R.~Kase, and S.~Tsujikawa,
\newblock Phys. Lett. {\bf B760}, 617 (2016), arXiv:1605.05565.

\bibitem{Kimura:2016rzw}
R.~Kimura, A.~Naruko, and D.~Yoshida,
\newblock JCAP {\bf 1701}, 002 (2017), arXiv:1608.07066.

\bibitem{Allys:2016kbq}
E.~Allys, P.~Peter, and Y.~Rodríguez,
\newblock Phys. Rev. {\bf D94}, 084041 (2016), arXiv:1609.05870.

\bibitem{Jimenez:2016upj}
J.~Beltrán~Jiménez and L.~Heisenberg,
\newblock Phys. Lett. {\bf B770}, 16 (2017), arXiv:1610.08960.

\bibitem{Emami:2016ldl}
R.~Emami, S.~Mukohyama, R.~Namba, and Y.-l. Zhang,
\newblock JCAP {\bf 1703}, 058 (2017), arXiv:1612.09581.

\bibitem{Maleknejad:2011jw}
A.~Maleknejad and M.~M. Sheikh-Jabbari,
\newblock Phys. Lett. {\bf B723}, 224 (2013), arXiv:1102.1513.

\bibitem{Maleknejad:2011sq}
A.~Maleknejad and M.~M. Sheikh-Jabbari,
\newblock Phys. Rev. {\bf D84}, 043515 (2011), arXiv:1102.1932.

\bibitem{Nieto:2016gnp}
C.~M. Nieto and Y.~Rodríguez,
\newblock Mod. Phys. Lett. {\bf A31}, 1640005 (2016), arXiv:1602.07197.

\bibitem{Oliveros:2019zkl}
A.~Oliveros and M.~A. Jaraba,
\newblock Int. J. Mod. Phys. {\bf D28}, 1950064 (2019), arXiv:1903.06005.

\bibitem{Rodriguez:2017wkg}
Y.~Rodríguez and A.~A. Navarro,
\newblock Phys. Dark Univ. {\bf 19}, 129 (2018), arXiv:1711.01935.

\bibitem{Kase:2018nwt}
R.~Kase and S.~Tsujikawa,
\newblock JCAP {\bf 1811}, 024 (2018), arXiv:1805.11919.

\bibitem{Heisenberg:2018mxx}
L.~Heisenberg, R.~Kase, and S.~Tsujikawa,
\newblock Phys. Rev. {\bf D98}, 024038 (2018), arXiv:1805.01066.

\bibitem{Kase:2018iwp}
R.~Kase and S.~Tsujikawa,
\newblock Phys. Rev. {\bf D97}, 103501 (2018), arXiv:1802.02728.

\bibitem{DeFelice:2016yws}
A.~De~Felice {\em et~al.},
\newblock JCAP {\bf 1606}, 048 (2016), arXiv:1603.05806.

\bibitem{DeFelice:2016uil}
A.~De~Felice {\em et~al.},
\newblock Phys. Rev. {\bf D94}, 044024 (2016), arXiv:1605.05066.

\bibitem{Nakamura:2017dnf}
S.~Nakamura, R.~Kase, and S.~Tsujikawa,
\newblock Phys. Rev. {\bf D95}, 104001 (2017), arXiv:1702.08610.

\bibitem{Rahman:2018fgy}
M.~Rahman and A.~A. Sen,
\newblock Phys. Rev. {\bf D99}, 024052 (2019), arXiv:1810.09200.

\bibitem{Heisenberg:2017hwb}
L.~Heisenberg, R.~Kase, M.~Minamitsuji, and S.~Tsujikawa,
\newblock JCAP {\bf 1708}, 024 (2017), arXiv:1706.05115.

\bibitem{Babichev:2016rlq}
E.~Babichev, C.~Charmousis, and A.~Lehébel,
\newblock Class. Quant. Grav. {\bf 33}, 154002 (2016), arXiv:1604.06402.

\bibitem{Chagoya:2017fyl}
J.~Chagoya, G.~Niz, and G.~Tasinato,
\newblock Class. Quant. Grav. {\bf 34}, 165002 (2017), arXiv:1703.09555.

\bibitem{Kase:2018owh}
R.~Kase, M.~Minamitsuji, and S.~Tsujikawa,
\newblock Phys. Lett. {\bf B782}, 541 (2018), arXiv:1803.06335.

\bibitem{Kase:2017egk}
R.~Kase, M.~Minamitsuji, and S.~Tsujikawa,
\newblock Phys. Rev. {\bf D97}, 084009 (2018), arXiv:1711.08713.

\bibitem{ErrastiDiez:2019trb}
V.~Errasti~Díez, B.~Gording, J.~A. Méndez-Zavaleta, and A.~Schmidt-May,
\newblock (2019), arXiv:1905.06968.

\bibitem{ErrastiDiez:2019ttn}
V.~Errasti~Díez, B.~Gording, J.~A. Méndez-Zavaleta, and A.~Schmidt-May,
\newblock (2019), arXiv:1905.06967.

\bibitem{Gomez:2019tbj}
L.~G. Gómez and Y.~Rodríguez,
\newblock (2019), arXiv:1907.07961.

\bibitem{Jimenez:2013qsa}
J.~Beltrán~Jiménez, R.~Durrer, L.~Heisenberg, and M.~Thorsrud,
\newblock JCAP {\bf 1310}, 064 (2013), arXiv:1308.1867.

\bibitem{Fleury:2014qfa}
P.~Fleury, J.~P. Beltrán~Almeida, C.~Pitrou, and J.-P. Uzan,
\newblock JCAP {\bf 1411}, 043 (2014), arXiv:1406.6254.

\bibitem{Hull:2015uwa}
M.~Hull, K.~Koyama, and G.~Tasinato,
\newblock Phys. Rev. {\bf D93}, 064012 (2016), arXiv:1510.07029.

\bibitem{deRham:2011by}
C.~de~Rham and L.~Heisenberg,
\newblock Phys. Rev. {\bf D84}, 043503 (2011), arXiv:1106.3312.

\bibitem{alexyeinzon2}
A.~Gallego~Cadavid and Y.~Rodríguez,
\newblock work in progress  (2019).

\end{thebibliography}
\end{document}